# Multidimensional quantum information based on single-photon temporal wavepackets

**Alex Hayat[†,] Xingxing Xing[†]*, Amir Feizpour, and Aephraim M. Steinberg**

*Centre for Quantum Information and Quantum Control, and Institute for Optical Sciences, Department of Physics, University of Toronto, 60 St. George Street, Toronto ON, M5S 1A7, Canada*
*\*xingxing@physics.utoronto.ca*

**Abstract:** We propose a multidimensional quantum information encoding approach based on temporal modulation of single photons, where the Hilbert space can be spanned by an in-principle infinite set of orthonormal temporal profiles. We analyze two specific realizations of such modulation schemes, and show that error rate per symbol can be smaller than 1% for practical implementations. Temporal modulation may enable multidimensional quantum communication over the existing fiber optical infrastructure, as well as provide an avenue for probing high-dimensional entanglement approaching the continuous limit.

**OCIS codes:** (270.5585) Quantum information and processing; (270.5565) Quantum communications.

**References and links:**

1. J. T. Barreiro, T.-C. Wei, and P. G. Kwiat, "Beating the channel capacity limit for linear photonic superdense coding," Nature Phys. **4**, 282-286 (2008).
2. J. T. Barreiro, N. K. Langford, N. A. Peters, and P. G. Kwiat, "Generation of Hyperentangled Photon Pairs," Phys. Rev. Lett. **95**, 260501 (2005).
3. T.Vértesi, S.Pironio, and N. Brunner, "Closing the Detection Loophole in Bell Experiments Using Qudits," Phys. Rev. Lett. **104**, 060401 (2010).
4. J. G. Rarity, and P. R. Tapster, "Experimental violation of Bell's inequality based on phase and momentum," Phys. Rev. Lett.**64**, 2495-2498 (1990).
5. J.D. Franson, "Bell inequality for position and time," Phys. Rev. Lett. **62**, 2205-2208 (1989).
6. Z. Y. Ou, X. Y. Zou, L. J. Wang, and L. Mandel, "Observation of nonlocal interference in separated photon channels," Phys. Rev. Lett. **65**, 321-324 (1990).
7. P. G. Kwiat, W. A. Vareka, C. K. Hong, H. Nathel, and R. Y. Chiao, "Correlated two-photon interference in a dual-beam Michelson interferometer," Phys. Rev. A **41**, 2910-2913 (1990).
8. A.Mair, A. Vaziri, G. Weihs, and A. Zeilinger, "Entanglement of the orbital angular momentum states of photons," Nature **412**, 313-316 (2001).
9. G. Molina-Terriza, J. P. Torres, and L. Torner, "Twisted photons," Nature Phys. **3**, 305-310 (2007).
10. A. C. Dada, J. Leach, G. S. Buller, M. J. Padgett, E. Andersson, "Experimental high-dimensional two-photon entanglement and violations of generalized Bell inequalities," Nature Phys. **7**, 677-680 (2011).
11. M. Halder, A. Beveratos, N. Gisin, V. Scarani, C. Simon, and H. Zbinden, "Entangling independent photons by time measurement," Nature Phys. **3**, 692-695 (2007).
12. J. Brendel, N. Gisin, W. Tittel, and H. Zbinden, "Pulsed Energy-Time Entangled Twin-Photon Source for Quantum Communication," Phys. Rev. Lett. **82**, 2594-2597 (1999).
13. R. T. Thew, A. Acín, H. Zbinden, and N. Gisin, "Bell-Type Test of Energy-Time Entangled Qutrits," Phys. Rev. Lett. **93**, 010503 (2004).
14. O.Kuzucu, F. N. C. Wong, S.Kurimura, and S.Tovstonog, "Joint temporal density measurements for two-photon state characterization," Phys. Rev. Lett. **101**, 153602 (2008).
15. A. Eckstein, B. Brecht, and C. Silberhorn, "A quantum pulse gate based on spectrally engineered sum frequency generation," Opt. Express **19**, 13770-13778 (2011).
16. B. Brecht, A. Eckstein, A. Christ, H. Suche, and C. Silberhorn, "From quantum pulse gate to quantum pulse shaper—engineered frequency conversion in nonlinear optical waveguides", New J. Phys. **13**,065029 (2011).

---

[†] *These authors contributed equally to this work.*

17. C. Polycarpou, K. N. Cassemiro, G. Venturi, A. Zavatta, and M. Bellini, "Adaptive Detection of Arbitrarily Shaped Ultrashort Quantum Light States", Phys. Rev. Lett. **109**, 053602 (2012).
18. Z. Y. Ou, and Y. J. Lu, "Cavity Enhanced Spontaneous Parametric Down-Conversion for the Prolongation of Correlation Time between Conjugate Photons", Phys. Rev. Lett.**83**, 2556–2559 (2000).
19. C. E. Kuklewicz, F. N. C.Wong, and J. H. Shapiro, "Time-bin-modulated biphotons from cavity-enhanced downconversion," Phys. Rev. Lett. **97**, 223601 (2006).
20. F. Wolfgramm, X. Xing, A. Cerè, A. Predojević, A. M. Steinberg, and M. W. Mitchell, "Bright filter-free source of indistinguishable photon pairs", Opt. Express **16**, 18145-18151 (2008).
21. D. Kielpinski, J. F. Corney, and H. M. Wiseman, "Quantum optical waveform conversion," Phys. Rev. Lett. **106**, 130501 (2011).
22. Y. Park, T.-J.Ahn, Y. Dai, J. Yao, and J.Azaña, "All-optical temporal integration of ultrafast pulse waveforms," Opt. Express **16**, 17817-17825 (2008).
23. J. L.Walsh, "A Closed Set of Normal Orthogonal Functions," Amer. J. Math. **45**, 5-24, (1923).
24. G. Golub and C. VanLoan, *"Matrix Computations"*, Johns Hopkins University Press,. Baltimore, $3^{rd}$ *ed.* (1996).
25. M. A. Nielsen, and I. L. Chuang, "Quantum Computation and Quantum Information" (Cambridge University, Cambridge, England, 2000).
26. G. Smith, "Quantum Channel Capacities," arXiv:1007.2855 (2010).
27. R. T. Thew, K. Nemoto, A. G. White, and W. J. Munro, "Qudit quantum-state tomography," Phys. Rev. A **66**, 012303 (2002).
28. C. K. Law, I. A. Walmsley and J. H. Eberly, "Continuous frequency entanglement: effective finite Hilbert space and entropy control," Phys. Rev. Lett. **84**, 5304 (2000).

1. **Introduction**

Multidimensional quantum information processing has been shown to open a wide range of possibilities including increased quantum channel capacity [1], multidimensional entanglement [2], and loophole-free Bell Inequality tests [3]. In contrast to the internal polarization degree of freedom of a photon, external degrees of freedom related to space [4] and time [5-7] are described by infinite-dimensional Hilbert spaces. The spatial degree of freedom has recently been employed to encode multidimensional quantum information using photon orbital angular momentum [8-10]; this approach, however, is not suitable for the single-mode fiber-optical communication infrastructure. The temporal degree of freedom has also been employed for both quantum communication and entanglement by encoding qubits in time bins [11]. Nevertheless, in these Franson-type unbalanced-interferometer approaches the dimension of the Hilbert space is not flexible because of discrete time bins determined once and for all by the structure of the interferometers [12, 13]. There have been recent proposals to use ultrafast nonlinear optical processes for projections onto continuous modes [14-17]. However, such broadband frequency-based approaches are impossible to implement in fiber-optical realizations due to the strong dispersion and nonlinearities in fibers. Lately, single-photon states with extremely long coherence times have become available through the use of cavity parametric down-conversion (PDC) [18-20], allowing manipulation of the temporal wavepacket of the photon using available electro-optics. Furthermore, long-coherence single-photon temporal compression and waveform conversion methods have been proposed for high-rate fiber telecommunications [21].

2. **Photon temporal-profile qudits**

Here we propose a multidimensional quantum information scheme, based on temporal modulation of single photons, and study its application in quantum communication and characterization of multidimensional time entanglement, approaching the continuous limit. In this scheme, the multidimensional Hilbert space basis states are, ideally, an infinite set of orthonormal temporal profiles. We develop a theoretical framework for the encoding and decoding of multidimensional quantum information based on temporal modulation and spectral filtering of photon wavepackets, and show that high efficiency and relatively low error rates can be achieved with commercially available technology. We propose specific realizations of the scheme, where the projection onto temporal modes is implemented by an electro-optical modulator (EOM) and a narrow-band optical filter. Practical EOMs have a finite bandwidth limiting the modulation speed, whereas the PDC-generated photons have an inherent bandwidth which determines the optimal width of the narrowband filter which can be employed. We show that the ratio of EOM modulation bandwidth to the bandwidth of the single photon limits the error rate per symbol (ERS) of such implementations, and we calculate the dependence of the ERS on encoding dimension and modulation speed.

If the single photon is produced in a narrow-band source (e.g. cavity-enhanced PDC), it can have a very long coherence time which makes it possible to achieve electro-optically modulated photon wavepackets in order to produce a variety of quantum states. As shown in Fig. 1a, the EOM modulation prepares the photon in a temporal profile $f_k(t)$, which we call a "symbol". The photon state can then be described as

$$|\psi_k\rangle = \int d\omega \, \tilde{f}_k(\omega) a^\dagger_\omega |0\rangle \qquad (1)$$

where $\tilde{f}_k(\omega)$ is the Fourier transform of $f_k(t)$. The overlap of two such states can be written in terms of the mode functions as follows:

$$p_{kj} = \left|\langle \psi_j | \psi_k \rangle\right|^2 = \left|\int d\omega' \tilde{f}_j^*(\omega') \int d\omega \tilde{f}_k(\omega) \langle 0 | a_{\omega'} a^\dagger_\omega | 0 \rangle\right|^2 = \left|\int dt f_j^*(t) f_k(t)\right|^2 \qquad (2)$$

For an orthonormal set of modes $f_j(t)$, $p_{kj} = \delta_{kj}$, where $\delta_{kj}$ is the Kronecker delta function.

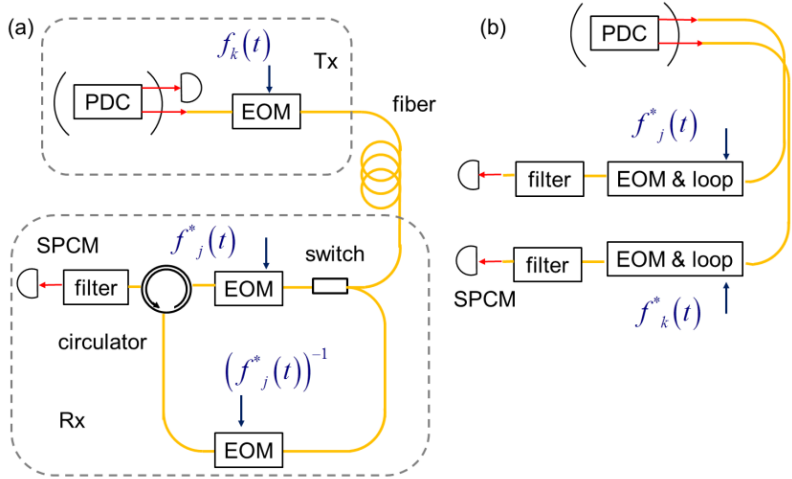

Fig. 1. (a) Temporal-pattern multidimensional quantum communication channel. In the transmitter (Tx) a temporal pattern $f_k(t)$ is written onto a triggered single photon. In the receiver (Rx) the photon is stored into a loop, where a different pattern is written on each round trip until the photon is transmitted through the filter and detected by a single photon counting module (SPCM), which is time-tagged for coincidences with the triggering SPCM in the Tx. (b) multidimensional time-entanglement state tomography scheme based on temporal pattern writing.

We are proposing a method to measure the photon in a given basis by using an EOM in a storage loop followed by a spectral filter (see Fig. 1a). In the storage loop, on round-trip $j$, the photon is modulated by a pattern $f_j^*(t)\left(f_{j-1}^*(t)\right)^{-1}$, such that after $j$ round trips it has acquired the mode structure $f_j^*(t)$, and the photon state can then be written as

$$U_j |\psi_k\rangle = \int d\omega \left[ \int d\Omega \tilde{f}_j^*(\Omega) \tilde{f}_k(\omega+\Omega) \right] a_\omega^\dagger |0\rangle ,$$ where the integral in the square bracket is the convolution of the two phase profiles in the frequency domain. The transmission of this state through the filter is given by the spectral overlap of the filter function $\tilde{T}(\omega)$ and the spectrum of $U_j |\psi_k\rangle$:

$$\overline{p}_{kj} = \int d\omega \left| \left[ \int d\Omega \tilde{f}_j^*(\Omega) \tilde{f}_k(\omega+\Omega) \right] \tilde{T}(\omega) \right|^2 \tag{3}$$

which can be written in the time domain as

$$\overline{p}_{kj} = \int dt \left| \left[ \int d\tau f_j^*(t) f_k(t) \right] T(\tau-t) \right|^2 , \tag{4}$$

where $T(t) = \int \tilde{T}(\omega) e^{i\omega t} d\omega$ is the time response of the filter. We show in the following that Eq. 4 can be a good approximation to the ideal projection (Eq. 2) under achievable conditions.

In the limit of a sufficiently narrowband filter, the timescale of the filter response, $\tau_{filter}$, can be long compared to the timescale of the photon wavepacket. In this limit, $T(t)$ can be approximated as a constant, $T_0$. The detector integration time is assumed to be much longer than the temporal width of the modulation profile function $f_k(t)$. The integration in Eq.4 can then be carried out to give

$$\overline{p}_{kj} \approx T_0^2 \tau_{filter} \left| \int d\tau f_j^*(t) f_k(t) \right|^2 = T_0^2 \tau_{filter} \left| \langle \Psi_j | \Psi_k \rangle \right|^2 , \tag{5}$$

which is proportional to the ideal overlap described in Eq. (2), with the proportionality constant $T_0^2 \tau_{filter}$. Such optical field integration filters have been demonstrated in classical applications based on fiber gratings [22]. For $k \neq j$, the photon is reflected at the filter, and is sent back into the loop via a circulator.

The cycle is repeated $d$ times for a Hilbert space of dimension $d$, and the photon incident in state $|\psi_k\rangle$ will be detected after $k$ round trips. This procedure thus corresponds to a projection in the $d$-dimensional basis (Fig. 1a). Such projection measurement can be also extended to characterize the multidimensional entanglement in the time domain, as shown in Fig. 1b.

It is important to note that our proposed scheme is capable of encoding and decoding any superposition of quantum states in any basis. This important property of a quantum communication channel is inherent in the use of electro-optical modulation. Such modulation enables imprinting any temporal function. More specifically, for any orthogonal basis, $\{|\psi_k\rangle\}$ defined by a set of temporal functions, $\{f_k(t)\}$, electro-optical modulation enables writing any linear combination of the set $\{f_k(t)\}$, and therefore any superposition of orthogonal temporal functions (symbols).

Our time-domain qudit detection introduces some time delay in principle; however in practical implementations this delay does not affect the symbol rates. Cavity-downconversion single-photon sources have relatively low photon generation rates – around several thousands narrowband photons per second [20]. On the other hand, the length of a photon wavepacket in the available narrowband sources is below 100ns. Therefore, even for extremely high qudit dimensions of several hundreds, the delay introduced by our detection scheme is several orders of magnitude shorter than the average time between photon generation events. Therefore, our multidimensional encoding approach not only does not reduce the symbol rate, but rather provides a unique solution for exploiting limited rates of true single-photon sources for high information rate by enlarging the alphabet.

In practice, filters have finite bandwidths, introducing deviations from the ideal formula (Eq. 5), which show up as some photons reaching the detector after the wrong number of roundtrips ("inter-symbol interference"). Different encoding schemes may have very different inter-symbol interference. In the following, we compare two possible encoding scenarios, and investigate the optimal parameter settings for each case, in terms of ERS and channel capacity.

### 3. Phase-flip encoding scheme

One multidimensional phase profile scheme, common in classical CDMA systems, is based on phase-flip modulation (PFM). In this scheme the single-photon wavepacket is divided into $n$ time intervals which have equal intensity integrals, and a phase of $0$ or $\pi$ radians is written on each of the $n$ time sections (Fig. 2a). By assigning certain phase profile patterns to these equal intensity-integral intervals, one can have a zero overlap integral (Eq. 5) for orthogonal phase profiles. A systematic method for the selection of these orthogonal states is implemented by using columns or rows of the Walsh matrix [23] for encoding. An $n$-by-$n$ Walsh matrix with $n>2$ exists if $n$ is a multiple of 4. Such a phase profile can be written as:

$$\phi_i(t) = \sum_{j=1}^{n} \frac{\pi}{2}(W_{ij}+1)\left[\Theta(t-t_j) - \Theta(t-t_{j+1})\right], i \in [1,n] \quad (6)$$

where $W_{ij}$ is a matrix element of the Walsh matrix, $\Theta(t)$ is a step function, and $t_j$ are the times that satisfy the following integral

$$\int_{-\infty}^{+\infty} \left(\Theta(t-t_j) - \Theta(t-t_{j+1})\right) \cdot |f(t)|^2 \, dt = \frac{1}{n} \int_{-\infty}^{+\infty} |f(t)|^2 \, dt. \quad (7)$$

In order to characterize the performance of this encoding scheme, we calculate the error rate per symbol (ERS), defined as the probability of incorrect projections normalized to the total detection rate, averaged over all possible inputs:

$$ERS = \frac{1}{d} \sum_{i=1}^{d} \frac{\sum_{j \neq i}^{d} \bar{p}_{ij}}{\sum_{j=1}^{d} \bar{p}_{ij}} = \frac{1}{d} \sum_{i=1}^{d} \left(1 - \frac{\bar{p}_{ii}}{\sum_{j=1}^{d} \bar{p}_{ij}}\right), \quad (8)$$

where $d$ is the encoding dimension and $\bar{p}_{ij}$ was defined in Eq. 5. The encoding dimension $d$ can be any number smaller than the Walsh Matrix dimension $n$. The ERS depends on the dimension of the Walsh matrix, since the larger the Walsh matrix, the larger the number of phase flips applied on average to the single photon. This leads to more high-frequency components in the photon spectrum, therefore a smaller spectral overlap with the narrowband filter (which is chosen to match the un-modulated photon bandwidth, as shown in Fig. 2b). For infinitely fast EOMs, we would wish to choose the dimension of the Walsh matrix as large as possible. However, for a given EOM speed, the largest possible Walsh matrix dimension is such that the fastest feature of the phase profile is slower than the EOM speed. To characterize the performance of the scheme, we wish to choose an optimal set of $d$ vectors within the *n*-dimensional vector space, and calculate its average ERS. In general, this is a computationally expensive optimization process. Instead, we observe that ERS is well correlated with the number of flips in the phase of the photon, and we will choose *d* phase profiles such that the average number of phase flips in products of all possible pairs of profiles (which correspond to the temporal profile resulting from encoding and decoding steps combined) is maximized. We define a dimensionless quantity *N* to characterize the EOM speed:

$$N = \Delta_{EOM} / \Delta_{photon}, \qquad (9)$$

where $\Delta_{EOM}$ and $\Delta_{photon}$ are the EOM bandwidth and photon linewidth, respectively. In Fig. 2c, we calculate the ERS as a function of encoding dimension *d*, for different Walsh matrix dimensions (*n*=16, 32, 64), for *N*=100. In this regime, the EOM speed is fast compared to the phase profile time scales, and the ERS decreases when we increase the Walsh matrix dimension. In this case, we have the ERS less than 0.06 for encoding dimensions up to 20. In Fig. 2d, we study the opposite case, taking *N*=20. Now the phase profiles have features faster than the EOM for Walsh matrix dimensions *n*=32 and 64, causing the optimization procedure described to deviate from the actual optimal set of vectors. In the limit of *n>>N*, the ERS approaches the limit *(d-1)/d* when all the symbols are equally likely to be detected independent of input symbols. As shown in Fig. 2d, we find that for encoding dimensions up to 9, the encoding scheme with n=64 performs worse than that with *n*=32 due to the slow EOM. For this specific EOM speed *N*=20, we obtain ERS of 0.48 for encoding dimension of 8.

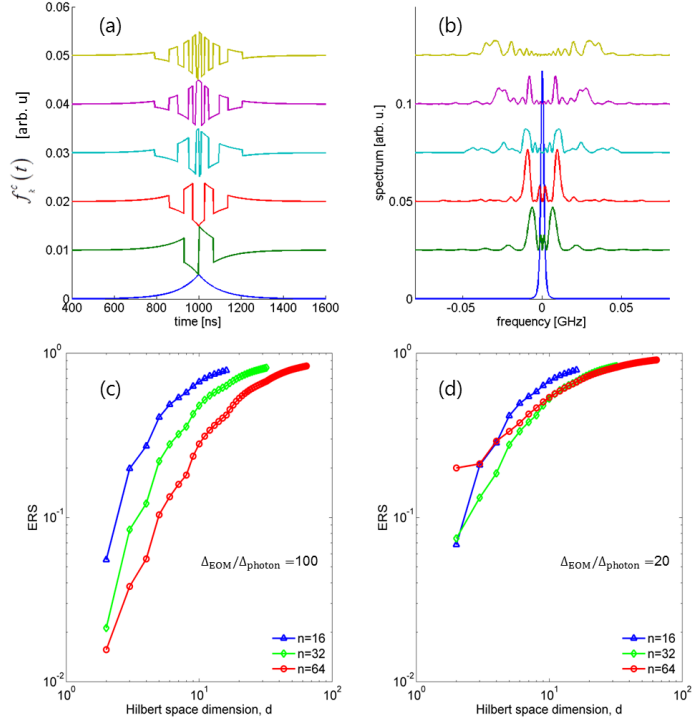

Fig. 2. (a) examples of phase-flip modulated photon profiles for a 100ns photon wavepacket– much slower than the modulation bandwidth for encoding dimension *d*=16 (b) Corresponding power spectra. Calculated ERS dependence on Hilbert space dimension for the phase-flip modulation scheme for different numbers of phase flips (c) for $\Delta_{EOM}/\Delta_{photon}$= 100 (d) for $\Delta_{EOM}/\Delta_{photon}$ = 20.

Even though, practically, such high ERS is not desirable, this provides a guideline to choose the optimal Walsh matrix dimension, that is, we should always choose the Walsh matrix dimension that minimizes the ERS. In this case the optimal Walsh matrix dimension, $n$, depends on the desired encoding dimension, $d$: for $d<10$, the optimal $n=32$, whereas for $d>10$ the optimal n=64. A finite speed of EOM, results also in slightly reduced efficiency of detection because the phase flip is not instantaneous, and thus even for a detection profile designed to undo an initial profile the photon spectrum will be modified and partially reflected by the filter.

### 4. Linear phase-ramp scheme

An alternative modulation scheme is linear phase modulation, where a linear temporal phase profile is applied to the single-photon pulse (Fig. 3 a), resulting in a frequency shift in the spectrum (Fig. 3 b),

$$|\psi_k\rangle = \int d\omega \tilde{f}_k(\omega - \Delta_k) a_\omega^\dagger |0\rangle \qquad (10)$$

If the frequency shift due to the linear phase pattern is large enough so that the modulated spectrum has a small overlap with the non-modulated one, the two states can be considered nearly orthogonal. For fast EOM modulations, our calculations show that this linear phase scheme results in ERS several orders of magnitude lower than that of the phase flip scheme. In practical EOM-based implementations, the speed of the EOM and the bandwidth of the photon can both affect the error rate in the detection. The encoding dimension $d$ is only limited by the dimensionless quantity $N$, $d \leq N$, as defined in Eq. 9. For a $d$-dimensional qudit, the linear phase rate $\Delta_k$ is chosen to be

$$\Delta_k = k \Delta_{EOM} / (d-1) \qquad (11)$$

for integer numbers $k > 0$ and $d \geq 2$. Some example of phase profiles for $d=16$ and their frequency spectra are given in Fig 3a and Fig 3b, respectively.

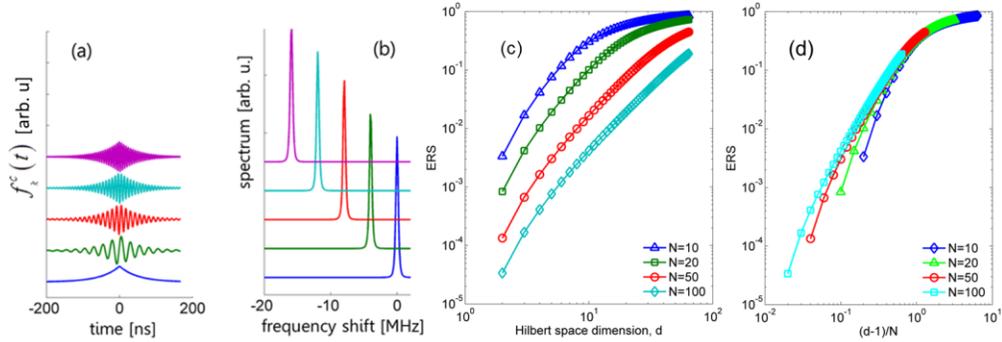

Fig 3. (a) The time dependence of the linear phase modulation profiles assuming a coherence time of 100ns for the photon and 1GHz modulation bandwidth. (b) Corresponding power spectra. (c) Calculated ERS dependence on Hilbert space dimension for the linear phase modulation scheme for different ratios of EOM bandwidth to photon bandwidth, $N=\Delta_{EOM}/\Delta_{photon}$. (d) Calculated ERS dependence on $(d-1)/N$ for different $N=\Delta_{EOM}/\Delta_{photon}$.

In Fig. 3c, we have plotted the ERS as a function of encoding dimension, and for illustration purposes, we have relaxed the requirement of $d \leq N$ and set $N=60$ for different EOM modulation bandwidth just to show the behavior of the scheme. For a constant $N$, the ERS increases with increasing encoding dimension, because of increasing overlap of the spectra of the two encoding profiles. The rate of the ERS increase with $d$ is larger for smaller $N$, which is a result of smaller spectral spacing between the different profiles. When the encoding dimension *d* approaches or exceeds *N*, the ERS saturates due to the fact that the spectra of the different profiles are already largely overlapping, and additional overlap does not affect the errors significantly. A faster EOM modulation speed leads to a smaller overlap, and thus a lower ERS. For example, with a 1ns modulation speed for a 100ns photon, the ERS can be made lower than 0.6% for encoding dimensions up to 10. The dominant factor of ERS is the spectral overlap of states with different phase profiles, which is determined by the spacing in frequency between two consecutive phase profiles. In Fig. 3d, we have shown the ERS as a function of the normalized dimension $d'=(d-1)/N$, which physically corresponds the number of symbols one encodes per available frequency bin determined by the EOM speed. For a given $d'$, the spacing in frequency is the same. The ERS for different EOM speed converges at $d' \sim 1$, as expected. For small $d'$, the ERS is slightly higher for higher dimensions. This is because for the same $d'$, the fixed frequency spacing leads to a fixed intersymbol interference rate for a given pair of symbols, but the larger number of symbols means the total ERS is higher.

## 5. Superpositions and mutual information

Ability to encode superpositions is a key requirement in quantum communication. Our scheme is capable of encoding and decoding any superpositions in any basis. This ability is inherent in the use of electro-optical modulation which can write any temporal function, and therefore in particular any superposition of orthogonal temporal functions (symbols). In fact, the ability to detect superpositions is the reason for the use of an electro-optical modulation decoding scheme, in contrast to frequency filtering only. This scheme, for example in a frequency basis, is capable of distinguishing between superposition states such as $|\Psi_+\rangle = 1/\sqrt{2}(|\omega_k\rangle + |\omega_j\rangle)$ and $|\Psi_-\rangle = 1/\sqrt{2}(|\omega_k\rangle - |\omega_j\rangle)$. A simple frequency filtering used in frequency division multiplexing is only capable of distinguishing between different frequencies $|\omega_k\rangle, |\omega_j\rangle$, however it would be incapable of detecting the phase between the two frequencies using simple filtering and to distinguish between $|\Psi_+\rangle = 1/\sqrt{2}(|\omega_k\rangle + |\omega_j\rangle)$ and $|\Psi_-\rangle = 1/\sqrt{2}(|\omega_k\rangle - |\omega_j\rangle)$.

In our scheme for any given computational basis $\{|C_k\rangle\}$ of dimension *d* with a corresponding set of temporal profiles $\{f_k^c(t)\}$ (Fig. 4 a), a different basis $\{|S_k\rangle\}$ can be constructed from superpositions of quantum states in $\{|C_k\rangle\}$. Similar to the case in the computational basis, superposition basis $\{|S_k\rangle\}$ resulting from the linear combinations of $\{|C_k\rangle\}$, corresponds to another set of temporal profiles $\{f_k^s(t)\}$, which can be also written by EOMs.

As a concrete example, we consider a linear-ramp encoding scheme (Section 4), for Hilbert space dimension of *d*=4. To generate a superposition basis, we start with a *d*×*d* matrix whose entries are random numbers such that the rows or columns of the matrix form a linearly independent set of vectors. With the Gram–Schmidt process [24] we orthonormalize the set to get a superposition basis $\{|S_k\rangle\}$, different from $\{|C_k\rangle\}$. A specific superposition basis $\{|S_k\rangle\}$ (Fig. 4 b), is constructed as described above:

$$|S_1\rangle = 0.267|C_1\rangle + 0.413|C_2\rangle - 0.785|C_3\rangle - 0.376|C_4\rangle$$
$$|S_2\rangle = 0.187|C_1\rangle + 0.593|C_2\rangle + 0.610|C_3\rangle - 0.491|C_4\rangle, \quad (12)$$
$$|S_3\rangle = 0.936|C_1\rangle - 0.324|C_2\rangle + 0.103|C_3\rangle + 0.094|C_4\rangle$$
$$|S_4\rangle = 0.134|C_1\rangle + 0.611|C_2\rangle - 0.006|C_3\rangle + 0.780|C_4\rangle$$

We calculate the matrix $\bar{p}_{kj}$, yielding the detection efficiencies and the error probabilities for encoding and decoding in both the computational basis $\{|C_k\rangle\}$ (Fig. 4 c), and in the superposition basis $\{|S_k\rangle\}$ using our scheme (Fig. 4 d). The diagonal entries of $\bar{p}_{kj}$ (Fig. 4 c, d) represent the probability of detection when the decoding profiles match the encoded ones ($k=j$), whereas the off-diagonal entries represent error probabilities. The superposition basis $\{|S_k\rangle\}$ shows slightly lower detection efficiency, however both $\{|S_k\rangle\}$ and $\{|C_k\rangle\}$ exhibit relatively high detection efficiency and very small error probabilities.

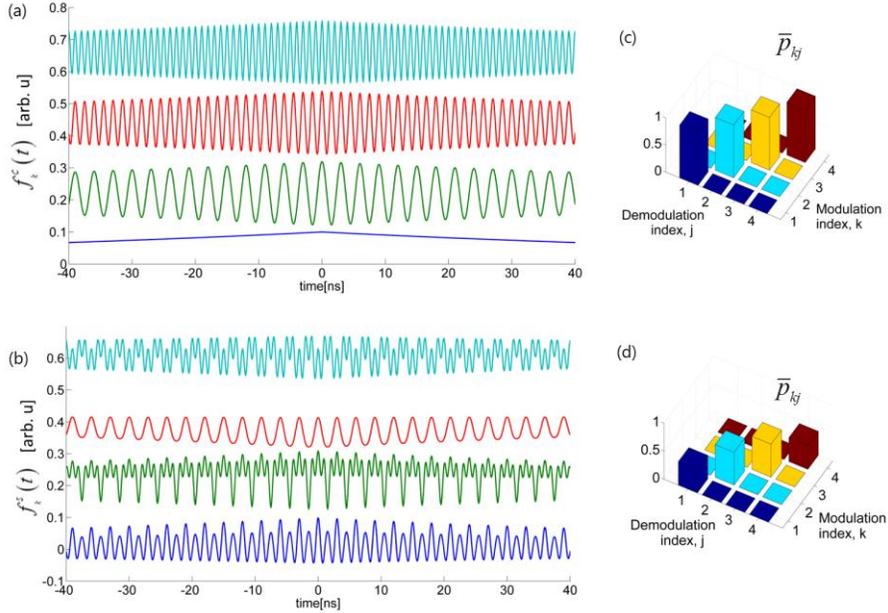

Fig. 4. (a) Temporal profiles of the $d=4$ linear ramp computational basis $\{|C_k\rangle\}$. (b) Temporal profiles of the superposition basis $\{|S_k\rangle\}$. Detection probability matrix $\bar{p}_{kj}$ (c) for the computational basis $\{|C_k\rangle\}$ and (d) for the superposition basis $\{|S_k\rangle\}$. The filter and EOM parameters are chosen $\Delta_{EOM}/\Delta_{photon}= 10$ and $\Delta_{filter}/\Delta_{photon} =100$.

Comparing the two encoding schemes proposed in sections 3 and 4, the linear phase encoding has several advantages relative to the phase flip scheme for multidimensional encoding, including higher efficiency of detection and lower error rates. As shown in Fig. 2, the ERS of phase-flip scheme is much higher than that of the linear phase ramp (Fig. 3), when other parameters are held constant. For this reason, in our following calculation of the mutual information, we will only consider the linear phase scheme. For ideal modulation and demodulation, the detection efficiency is unity and the ERS is zero, and there is no information loss over the channel. Practically, however, the information can be lost due to less-than-unity detection efficiency and finite ERS, and there is a trade-off between the detection efficiency and ERS when choosing the filter bandwidth. To characterize the quantum channel, one can calculate the mutual information between the input and the output for such a channel.

The mutual information measures the amount of information transmitted over a quantum channel. It is defined as $I(X:Y) \equiv H(Y) - H(Y|X)$; for the input variable X and output variable Y, where H(Y) is the entropy of the variable Y, and H(Y|X) is the conditional entropy of Y, given X. The calculation of mutual information $I(X:Y)$ [25] between the input X and output Y is shown in Fig. 5a. To obtain channel capacity one has to perform optimization over all possible input states, which is difficult in general [26]. Here we have limited our choice of parameters to the states of the form

$$|\Psi\rangle = \sqrt{1-a^2}|\omega_k\rangle + ae^{ib}|\omega_j\rangle \qquad (13)$$

where *a* and *b* are the amplitude and phase parameters, respectively. In order to project the photon state onto $|\Psi\rangle$, in general we need to introduce amplitude modulation, since such superposition in the frequency domain means an intensity beating pattern in the time domain, which necessarily leads to loss if we use only passive optical elements such as EOMs, with no gain medium. Such loss is basis dependent, and when the superposition has equal amplitudes of different frequencies, the loss is maximized to 50%. However, this is not a limitation of the scheme, since such losses can be calibrated ahead of time for the demodulation basis used. There is also a trade-off between transmission and ERS while varying the filter bandwidth. We can determine the optimum filter bandwidth using the calculated mutual information curve.

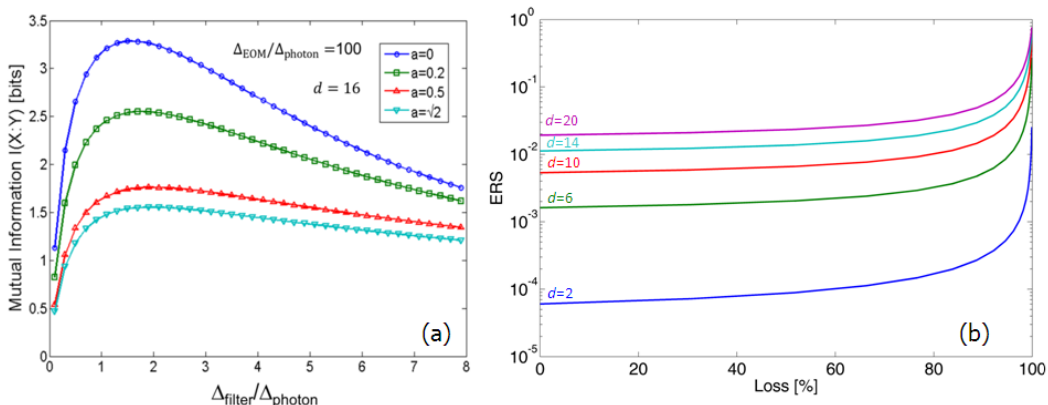

Fig 5. (a) Mutual information I(X:Y) versus the ratio of filter bandwidth to photon linewidth, $\Delta_{filter}/\Delta_{photon}$. for various values of superposition parameter a, as defined in Eq. 11. The modulation speed of the EOM is chosen so that $\Delta_{EOM}/\Delta_{photon} = 100$. The calculation is done for the linear phase ramp scheme, with the encoding dimension *d*=16 (for this dimensionality the ideal information is 4 bits/symbol). (b) Dependence of the ERS on the loss level for for different Hilbert space dimensions, *d*, with dark count rate of 100/s, and 100ns photon coherence time.

As shown in Fig. 5a, for encoding dimension *d*=16, the optimal filter bandwidth $\Delta_{filter}$ is found to be approximately $1.5\Delta_{photon;}$ for this value, the mutual information is 3.3 bits/symbol, while for equal-amplitude superpositions it drops considerably, but is still well above 1.

State-independent loss in absence of dark counts in the detectors, does not affect the ERS because both the detection efficiency and the error rate are reduced with increasing loss equally. Dark counts in the detectors, however, can affect the ERS. We calculate the effect of loss on the ERS for various encoding dimensions for experimentally available sources of single photons with 100ns coherence time, and detectors with dark count rate of 100/s (Fig. 5 b). The ERS is significantly increased for loss levels higher than 80%; and for higher Hilbert space dimensions *d* the loss has a stronger effect on ERS. For very high loss levels, the ERS reaches a maximum value of 1-1/*d*. Nevertheless, at loss levels below 60%, the loss-induced ERS is comparable to that caused by the inter-symbol interference, so that the overall ERS remains low.

The detection scheme described above can be used for characterization of high-dimensional entanglement in the time domain, using the techniques previously developed for orbital-angular-momentum qudits [27]. In contrast to the time-bin approach [11, 12], where the number of time bins are limited to the number of modes in the spatial interferometer, in our scheme a projection in the detector (Fig. 1b) involves interference between many different time bins. In principle this time-entanglement characterization can be nearly continuous; however in practice, the time resolution of the scheme is limited by the speed of the EOM relative to the coherence length of the photon. Law *et. al.* have shown, based on the Schmidt decomposition, that the effective Hilbert space of the PDC biphoton state can be finite-dimensional and therefore, it is possible to reconstruct the biphoton state with high fidelity using a finite-dimensional projection [28]. The number of basis vectors needed for proper mapping depends on the spectral width of the pump and the phase-matching profile of the crystal, as well as any other function that modifies the spectrum of the down-conversion. For example, in cavity-based PDC pumped by a CW laser, the number of basis functions *M* is given by the ratio of the cavity linewidth to that of the pump. In this case superpositions of *M* low-frequency modes are sufficient for tomography and reconstruction of the wavefunction with high fidelity. High error rates could prevent the violation of Bell's inequalities by reducing the two-photon interference visibility below the classical field theory limit [4, 6]. In our scheme the error rates depend on the specific realization, the photon bandwidth and on the modulator speed. For the linear ramp scheme, with $\Delta_{EOM}/\Delta_{photon} \sim 100$, even for Hilbert space dimensions as high as 10, the error rates are well below 0.6%, which is smaller than the error rates in previous experiments demonstrating violations of Bell's inequalities for qubits [4] and for high-dimensional qudits [10].

## 6. Conclusions

In conclusion, we have proposed a multidimensional quantum information scheme based on temporal modulation of long-coherence-time photons. We have developed a theoretical model for the multidimensional temporal profile encoding, and calculated the efficiency and error rate per symbol for two encoding schemes considering practical electro-optical modulators. Our results show that reasonably low error rate per symbol (smaller than 0.6% for dimensions up to 10) can be obtained, and the linear phase profile scheme is preferable to the (classically common) phase flip encoding scheme, due to imperfections resulting from the difference between spectral filtering and ideal Hilbert-space projections. The temporal encoding scheme may make it feasible to carry out multidimensional quantum information processing and quantum communication on the existing fiber optical telecommunication infrastructure, as well as probing multidimensional time entanglement approaching the limit of continuous-time measurements.

We are grateful to Daniel Gottesman and Debbie Leung for many helpful discussions about optimization in the phase flip encoding. We appreciate financial support from the Natural Sciences and Engineering Research Council (NSERC) of Canada and from the Canadian Institute for Advanced Research (CIFAR).